\documentclass{iopart}
\usepackage[dvips]{graphicx}
\usepackage{amssymb}
\usepackage{wasysym}
\newcommand{\eqref}[1]{(\ref{#1})}
\newcommand{\abs}[1]{\left|#1\right|}
\newcommand{\text}[1]{\hbox{\scriptsize\rm #1}}
\newcommand{\un}[1]{\mathrm{\,#1}}
\newcommand{\Qhat}{{\widehat{Q}}}
\newcommand{\What}{{\widehat{W}}}

\newcommand{\deltaf}{\delta\!f}

\newcommand{\PRpt}{{\it Phys. Rep.}}

\newcommand{\mod}{\mathop{\rm mod}\nolimits}
\newcommand{\Real}{\mathop{\rm Re}\nolimits}
\newcommand{\Imag}{\mathop{\rm Im}\nolimits}
\begin{document}
\title[Data Analysis for LIGO-ALLEGRO Stochastic Search]
{A Data Analysis Technique for the
  LIGO-ALLEGRO Stochastic Background Search}
\author{John T Whelan$^1$, Sukanta Bose$^2$, Jonathan Hanson$^3$,
  Ik Siong Heng$^4$, Warren W Johnson$^3$, Martin P McHugh$^1$
  and Peter Zhang$^3$}
\address{$^1$ Department of Physics, Loyola University, New Orleans,
  Louisiana 70118, USA}
\address{$^2$Washington State University, Pullman, Washington 99164, USA}
\address{$^3$ Department of Physics and Astronomy, Louisiana State
  University, Baton Rouge, Louisiana 70803, USA}
\address{$^4$ Department of Physics and Astronomy,
  University of Glasgow, Glasgow, G12 8QQ, United Kingdom}
\begin{abstract}
  We describe the cross-correlation measurements being carried out on
  data from the LIGO Livingston Observatory and the ALLEGRO resonant
  bar detector.  The LIGO data are sampled at 16384\,Hz while the
  ALLEGRO data are base-banded, i.e., heterodyned at 899\,Hz and then
  sampled at 250\,Hz.  We handle these different sampling parameters
  by working in the Fourier domain, and demonstrate the approximate
  equivalence of this measurement to a hypothetical time-domain method
  in which both data streams are upsampled.
\end{abstract}
\ead{jtwhelan@loyno.edu}

\section{Introduction}
\label{s:intro}

Analysis is currently underway to search for the signature of a
stochastic gravitational-wave background (SGWB) by correlating the
signals of the 4\,km interferometer (IFO) at the LIGO Livingston
Observatory (LLO)\cite{LLO,LIGOS1} with the ALLEGRO resonant bar
detector\cite{ALLEGRO,Mauceli:1996,Harry:1999}.  As described
elsewhere\cite{Whelan:2003}, the LLO-ALLEGRO experiment is sensitive
to a higher frequency range than the corresponding experiment using
LLO and the IFOs at the LIGO Hanford Observatory (LHO)\cite{StochS1},
thanks to the relative proximity of the ALLEGRO and LLO sites.
Additionally, the ALLEGRO detector can be rotated, changing the
response of the experiment to a SGWB and thus providing a means to
distinguish gravitational-wave (GW) correlations from correlated
noise.\cite{Finn:2001}.

The results of a cross-correlation measurement using data taken during
LIGO's second science run (S2) will be reported in the near
future.\cite{ALLEGROS1} The present work describes some details of the
analysis procedure used, notably the handling of the different
sampling rates of the data taken by the two detectors, and the
heterodyned nature of the ALLEGRO data.  The procedure applied here
may prove useful in coherent measurements involving data sampled at
different rates, such as data from the LIGO and Virgo\cite{Virgo}
IFOs.

\section{Stochastic Gravitational-Wave Backgrounds}
\label{s:stoch}

\subsection{Definitions}

A gravitational wave (GW) is described by the metric tensor
perturbation $h_{ab}(\vec{r},t)$.  A given GW detector, located at
position $\vec{r}_{\text{det}}$ on the Earth, will measure a GW strain
which is some projection of this tensor:
\begin{equation}
  h(t) = h_{ab}(\vec{r}_{\text{det}},t) d^{ab}
\end{equation}
where $d^{ab}$ is the detector response tensor, which is
\begin{equation}
  d_{\text{(ifo)}}^{ab} = \frac{1}{2}(\hat{x}^a \hat{x}^b - \hat{y}^a \hat{y}^b)
\end{equation}
for an interferometer with arms parallel to the unit vectors $\hat{x}$
and $\hat{y}$ and
\begin{equation}
  d_{\text{(bar)}}^{ab} = \hat{u}^a \hat{u}^b
\end{equation}
for a resonant bar with long axis parallel to the unit vector
$\hat{u}$.

\begin{figure}[htbp]
  \vspace{5pt}
  \begin{center}
    \includegraphics[height=3.5in]{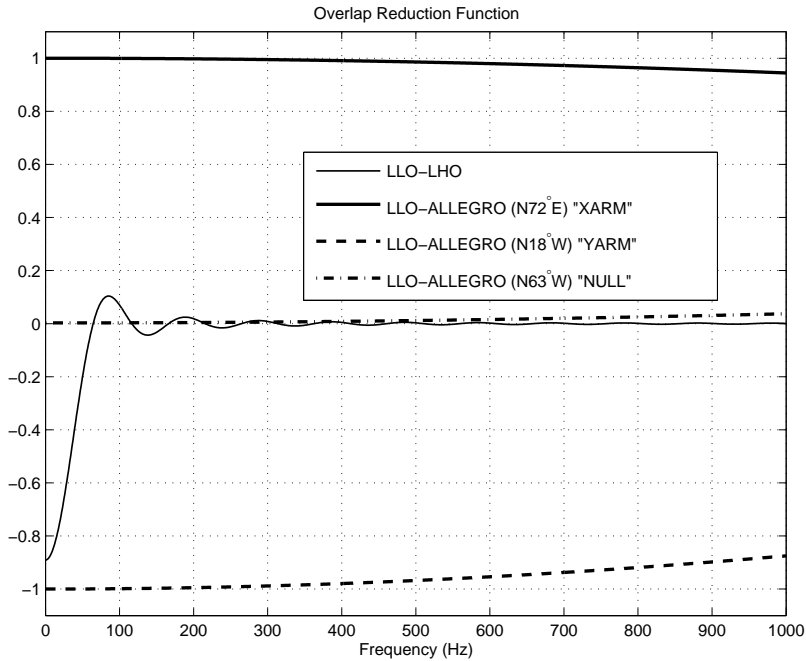}    
  \end{center}
  \caption{ The overlap reduction function for LIGO Livingston
    Observatory (LLO) with ALLEGRO and with LIGO Hanford Observatory
    (LHO).  The three LLO-ALLEGRO curves correspond to the three
    orientations in which ALLEGRO was operated during LIGO's S2 run:
    ``XARM'' (N72$^\circ$E) is nearly parallel to the x-arm of LLO
    (``aligned''); ``YARM'' (N18$^\circ$W) is nearly parallel to the
    y-arm of LLO (``anti-aligned''); ``NULL'' (N$63^\circ$W) is
    halfway in between these two orientations (a ``null alignment''
    midway between the two LLO arms).  The LLO-LHO overlap reduction
    function is shown for reference.}
  \label{fig:overlap}
\end{figure}

A stochastic GW background (SGWB) can arise from a superposition of
uncorrelated cosmological or astrophysical sources.  Such a
background, if isotropic, unpolarized, stationary and Gaussian, will
generate a cross-correlation between the strains measured by two
detectors.\cite{Christensen:1992,Allen:1997,Maggiore:2000} In terms of
the continuous Fourier transform defined by
\begin{equation}
  \label{eq:cft}
  \widetilde{a}(f)=\int_{-\infty}^\infty dt\,a(t)\,\exp(-i2\pi f[t-t_0])
  \ ,
\end{equation}
(where $t_0$ is an arbitrarily-chosen time origin) the expected
cross-correlation is
\begin{equation}
  \label{e:h1h2corr}
  \langle \widetilde{h}_1^*(f) \widetilde{h}_2(f') \rangle
  = \frac{1}{2} \delta(f-f')\,P_{\text{gw}}(f)\,\gamma_{12}(f)
\end{equation}
where
\begin{equation}
  \gamma_{12}(f)={d_{1ab}}\, {d_2^{cd}}\,
  \frac{5}{4\pi} \iint d^2\Omega_{\hat{n}}\,
  P^{\text{TT}\hat{n}}{}^{ab}_{cd}\,
  e^{i2\pi f\hat{n}\cdot(\vec{r}_2-\vec{r}_1)/c}
\end{equation}
is the overlap reduction function\cite{Flanagan:1993} between the two
detectors, defined in terms of the projector
$P^{\text{TT}\hat{n}}{}^{ab}_{cd}$ onto traceless symmetric tensors
transverse to the unit vector $\hat{n}$.
Figure~\ref{fig:overlap} shows the
overlap reduction functions for several detector pairs of interest.

$P_{\text{gw}}(f)$ is the one-sided spectrum of the SGWB.  This is the
one-sided power spectral density (PSD) the background would generate
in an interferometer with perpendicular arms, which can be seen from
\eqref{e:h1h2corr} and the fact that the overlap reduction function of
such an interferometer with itself is unity.  Since the overlap
reduction function of a resonant bar with itself is $4/3$ (see
\cite{bargeom} for more details), the PSD of
the strain measured by a bar detector due to the SGWB would be
$(4/3)P_{\text{gw}}(f)$.

A related measure of the spectrum is the dimensionless quantity
$\Omega_{\text{gw}}$, the GW energy density per unit logarithmic
frequency divided by the critical energy density $\rho_{\text{c}}$
need to close the universe:
\begin{equation}
  \label{e:Omega_gw}
  \Omega_{\text{gw}}(f)= \frac{f}{\rho_{\rm c}}\,
  \frac{d\rho_{\text{gw}}}{df}
  = \frac{10\pi^2}{3 H_0^2}\,f^{3} P_{\text{gw}}(f)
  \ .
\end{equation}

\subsection{Detection Method}

The standard method to search for such a background is to
cross-correlate the outputs of two gravitational wave detectors
\cite{Christensen:1992}.  If each detector signal $s_{1,2}(t)$
is assumed to be made up of a gravitational wave
component $h_{1,2}(t)$ plus an instrumental noise component
\begin{equation}
  s_{1,2}(t)=h_{1,2}(t)+n_{1,2}(t)
\end{equation}
and the noise in the two detectors is approximately
uncorrelated but significantly
larger than the gravitational-wave signal, then the average
cross-correlation should come from the stochastic GW background:
\begin{equation}
  \label{eq:s1s2corr}
  \langle \widetilde{s}_1^*(f) \widetilde{s}_2(f')\rangle
  \approx \langle \widetilde{h}_1^*(f) \widetilde{h}_2(f')\rangle
  = \frac{1}{2} \delta(f-f')\,P_{\text{gw}}(f)\,\gamma_{12}(f)
\end{equation}
while the average auto-correlation should come from the noise:
\begin{equation}
  \label{eq:s1s1corr}
  \langle \widetilde{s}_{1,2}^*(f) \widetilde{s}_{1,2}(f')\rangle
  \approx \langle \widetilde{n}_{1,2}^*(f) \widetilde{n}_{1,2}(f')\rangle
  = \frac{1}{2} \delta(f-f')\,P_{1,2}(f)
\end{equation}
If we construct a cross-correlation statistic
\begin{equation}
  Y^c
  =
  \int dt_1\, dt_2\, {s_1(t_1)}\, {Q(t_1-t_2)}
  {s_2(t_2)}
  \label{eq:CCstat}
  =
  \int df\,{\widetilde{s}_1^*(f)}\, {\widetilde{Q}(f)}\,
  {\widetilde{s}_2(f)}
\end{equation}                                
the expected statistics of \eqref{eq:CCstat} are given by\cite{Allen:1999}
\begin{equation}
  \label{eq:contmean}
  \mu_{Y^c} = \langle Y^c\rangle 
  \approx \frac{T}{2}
  \int_{-\infty}^{\infty} df \,\gamma(\abs{f})\,P_{\text{gw}}(f)\widetilde{Q}(f)
\end{equation}
and
\begin{equation}
  \label{eq:contvar}
  \sigma_{Y^c}^2 = \langle (Y^c-\mu_{Y^c})^2\rangle
  \approx \frac{T}{4}
  \int_{-\infty}^{\infty} df \,P_{1}(f)\,P_{2}(f)\abs{\widetilde{Q}(f)}^2
\end{equation}
Using \eqref{eq:contmean} and \eqref{eq:contvar}, the optimal choice
for the filter $\widetilde{Q}(f)$, given a predicted shape for the
spectrum $P_{\text{gw}}(f)$ can be shown\cite{Allen:1999} to be
\begin{equation}
  \label{eq:Qopt}
  \widetilde{Q}(f) \propto \frac{\gamma(\abs{f})\,P_{\text{gw}}(f)}
  {P_{1}(f)\,P_{2}(f)}
\end{equation}
This has negligible support when $P_{1}(f)$ or $P_{2}(f)$ is large,
which allows one to limit the integration in \eqref{eq:CCstat} to a
finite range of frequencies.

In Sec.~\ref{s:method} we consider how the approximate continuous-time
expressions \eqref{eq:contmean} and \eqref{eq:contvar} manifest
themselves given the different discrete sampling parameters of the
LIGO and ALLEGRO detectors.

\section{Data Analysis Technique}
\label{s:method}

\subsection{Frequency-Domain Method}

As described in \cite{Whelan:2003}, the frequency range of sensitivity
correlation measurements involving the ALLEGRO resonant bar detector
and the LIGO Livingston Observatory 40\,km away is determined by the
sensitive bandwidth of ALLEGRO \cite{ALLEGRO,Harry:1999}, which is
limited to a band a few tens of hertz wide near 900\,Hz.  Due to the
relatively narrow-band nature of the detector response, the data are
base-banded, i.e., heterodyned at 899\,Hz (during the S2 run) and
downsampled to 250\,Hz, so that the data represent a frequency range
from $(899-125)\un{Hz}=774\un{Hz}$ to $(899+125)\un{Hz}=1024\un{Hz}$.
The LIGO data are sampled at a frequency of 16384\,Hz and therefore
represent frequencies ranging from $-8192\un{Hz}$ to $8192\un{Hz}$.

Previous work\cite{Whelan:2003} proposed resampling and heterodyning the
LIGO data before cross-correlation with ALLEGRO data.  However, since
this would involve not only downsampling by powers of two but also
upsampling by a factor of $5^3$, the current approach is to
approximate \eqref{eq:CCstat} in the frequency domain.

First, we describe the relationship between the discretely sampled data
in each instrument and the underlying continuous time series $s_{1,2}(t)$.
The first time series, which we will take to be the LIGO data, is
sampled at a frequency $(\delta t_1)^{-1}$, which for the sake of this
example we will take to be $2048\un{Hz}$.\footnote{The actual sampling
  rate is 16384\,Hz, but we digitally downsample the data before
  analysis, a process which is straightforward enough that we don't
  need to explicitly address it here.} Before sampling, it is low-pass
filtered to avoid aliasing of higher-frequency data into the analysis
band.  We idealize the effects of this anti-aliasing filter by
defining
\begin{equation}
  \widetilde{S}_1(f)
  =
  \cases{
    \widetilde{s}_1(f)
    &
    $\abs{f} < \frac{1}{2\,\delta{t}_1}$ 
    \\
    0 & $\abs{f} > \frac{1}{2\,\delta{t}_1}$
  }
\end{equation}
and writing the discretely-sampled signal as
\begin{equation}
  S_1[j] = S_1(t_0+j\,\delta{t}_1) \qquad j=0,\ldots,N_1-1
\end{equation}
The ALLEGRO data are first heterodyned at $f^h_2=899\un{Hz}$ to
produce a continuous time series
\begin{equation}
  s_2^h(t) = e^{-i2\pi f^h_2(t-t_0)} s_2(t)
\end{equation}
then anti-alias filtered
\begin{equation}
  \widetilde{S}_2^h(f)
  =
  \cases{
    \widetilde{s}_2^h(f) = \widetilde{s}_2(f^h_2+f)
    &
    $\abs{f} < \frac{1}{2\,\delta{t}_2}$ 
    \\
    0 & $\abs{f} > \frac{1}{2\,\delta{t}_2}$
  }
\end{equation}
and sampled at $(\delta t_2)^{-1}=250\un{Hz}$ to obtain 
\begin{equation}
  S_2^h[k] = S_2^h(t_0+k\,\delta{t}_2) \qquad k=0,\ldots,N_2-1  
\end{equation}
Note that this signal is now intrinsically complex.  The
``unheterodyned'' time series $S_2(t)=e^{i2\pi f^h_2(t-t_0)} S_2^h(t)$
is a band-passed version of the original signal
\begin{equation}
  \widetilde{S}_2(f)
  =
  \cases{
    \widetilde{s}_2(f)
    &
    $\abs{f-f^h_2} < \frac{1}{2\,\delta{t}_2}$ 
    \\
    0 & $\abs{f-f^h_2} > \frac{1}{2\,\delta{t}_2}$
  }
\end{equation}
While $s_2(t)$ is real, $S_2(t)$ is not, as a result of the bandpass.

\begin{figure}[htbp]
  \vspace{5pt}
  \begin{center}
    \includegraphics[height=3in]{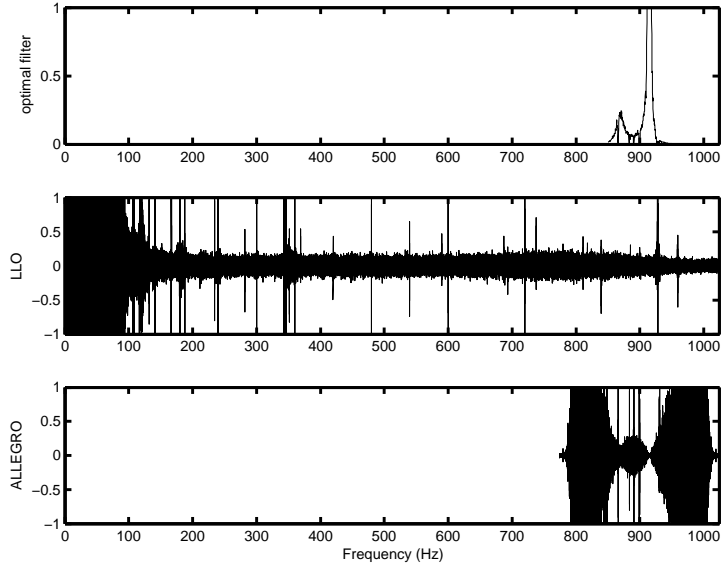}
  \end{center}
  \caption{The frequency ranges covered by Fourier-transformed
    LLO and ALLEGRO data and
    the optimal filter used for analysis.  The vertical scales are
    arbitrary and the data are meant to be illustrative only.  If LLO
    data are downsampled to 2048\,Hz, their discrete Fourier transform
    covers frequencies from -1024\,Hz to 1024\,Hz.  The ALLEGRO data
    are heterodyned at 899\,Hz and sampled at 250\,Hz, which makes the
    frequencies represented range from 774\,Hz to 1024\,Hz.  The
    optimal filter has non-negligible support only for
    $850\,Hz\lesssim f\lesssim 950\,Hz$, so those are the only
    frequencies included in the cross-correlation.}
  \label{fig:freqs}
\end{figure}

In our analysis, we construct an ensemble of statistics, each
calculated by cross-correlating $T=60\un{s}$ worth of data for each
instrument, which amounts to $N_1=T/(\delta t_1)=122880$ points worth
of LLO data and $N_2=T/(\delta t_2)=15000$ points worth of ALLEGRO
data.  The LLO data are windowed and zero-padded out to a length
$M_1=2N_1$ and discrete Fourier transformed to produce
\begin{equation}
  \widehat{wS}_1[\ell] = \sum_{j=0}^{N_1-1} w_1[j]S_1[j]e^{-i2\pi\ell j/M_1}
  = \sum_{j=0}^{N_1-1} w_1[j]S_1[j]e^{-i2\pi\ell\,\deltaf\,j\,\delta t_1}
\end{equation}
where $\delta f=(\delta t_1\,M_1)^{-1}=(2T)^{-1}=\frac{1}{120}\un{Hz}$
for $T=60\un{s}$, and $\ell=-N_1,\ldots,N_1-1$.
Meanwhile, the ALLEGRO data are windowed and
zero-padded out to a length $M_2=2N_2$ and discrete Fourier
transformed to produce
\begin{eqnarray}
  \widehat{wS}_2[\ell] &=& \sum_{k=0}^{N_2-1} w_2[k]S_2^h[k]e^{-i2\pi\ell k/M_2}
  e^{i2\pi f^h_2\,k\,\delta t_2}
  \nonumber
  \\
  &=& \sum_{k=0}^{N_2-1} w_2[k]S_2^h[k]
  e^{-i2\pi(\ell\,\deltaf-f^h_2)\,k\,\delta t_2}
\end{eqnarray}
where $\delta f=(\delta t_2\,M_2)^{-1}$ is the same frequency
resolution as before, and $\ell=\frac{f^h_2}{\delta
  f}-N_2,\ldots,\frac{f^h_2}{\delta f}+N_2-1$.  With these
definitions,
\begin{equation}
  \delta t_{1,2}\,\widehat{wS}_{1,2}[\ell]
  \sim\widetilde{S}_{1,2}(\ell\,\deltaf)
\end{equation}
so if we construct a frequency-domain optimal filter
$\widetilde{Q}(f)$, we can obtain an approximate analogue
$Y\sim Y^c$ for the
$Y^c$ defined in \eqref{eq:CCstat} by constructing the statistic
\begin{equation}
  \label{eq:discfreqstat}
  Y = \sum_{\ell=\ell_{\text{min}}}^{\ell_{\text{max}}}
  \deltaf \, (\delta t_1\,\widehat{wS}_1[\ell])^*\,
  \widetilde{Q}(\ell\,\deltaf)\,
  (\delta t_2\,\widehat{wS}_2[\ell])
\end{equation}
where $(\ell_{\text{min}}\,\deltaf,\ell_{\text{max}}\,\deltaf)$ is the
frequency range over which $\widetilde{Q}(f)$ has significant support.
Because of the factor of $P_1(f)P_2(f)$ in the denominator of
\eqref{eq:Qopt}, this is limited to a subset of the frequency ranges
associated with both the LLO and ALLEGRO data, as illustrated in
Figure~\ref{fig:freqs}.

\subsection{Time-Domain Equivalent}

In this section we consider a time-domain cross-correlation equivalent
to the discrete frequency-domain approximation \eqref{eq:discfreqstat}
and explicitly calculate the discrete-time equivalents of
\eqref{eq:contmean} and \eqref{eq:contvar}.

One conceptually simple approach to cross-correlating these two data
streams in the time domain would be to upsample both to the same
sampling frequency.  This should work as long as the ratio of the
sampling rates is a rational number.  We define $r_1$ and $r_2$ as the
smallest integers such that
\begin{equation}
  \frac{\delta{t}_1}{\delta{t}_2} = \frac{r_1}{r_2}
\end{equation}
(in the present example, $r_1=125$ and $r_2=1024$) and then define
\begin{equation}
  \delta{t} = \frac{\delta{t}_1}{r_1} = \frac{\delta{t}_2}{r_2}
\end{equation}
(In the present example, $(\delta t)^{-1}=256000\un{Hz}$.)
These upsampled time series would then each contain
\begin{equation}
  N = N_1 r_1 = N_2 r_2
\end{equation}
points.  (In the present example, $N=15360000$.)
The method used to upsample the data does not interest us,
since we won't actually perform the time-domain upsampling.  Instead,
we idealize the result as the discrete sampling of the band-passed
time series $S_1(t)$ and $S_2^h(t)$:
\begin{eqnarray}
  S_1^r[J] &=& S_1(t_0+j\,\delta{t}) \qquad J=0,\ldots,N \\
  S_2^{hr}[K] &=& S_2^h(t_0+K\,\delta{t}) \qquad K=0,\ldots,N
\end{eqnarray}

If we also assume that $f^h_2+1/2\delta{t}_2<1/2\delta{t}$, there is
no loss of information in going between the heterodyned and
non-heterodyned data streams at the higher sampling rate.  It is 
thus reasonable to think of the cross-correlation statistic as
\begin{eqnarray}
  \label{eq:Yrsum}
    Y^r &=& \sum_{J=0}^{N-1} \delta{t} \sum_{K=0}^{N-1} \delta{t}
    \,w_1^r[J]\,S_1^r[J]^*\,Q^r[J-K]\,
    w_2^r[K]\,S_2^r[K]
    \\
    \nonumber
    &=& \sum_{J=0}^{N-1} \delta{t} \sum_{K=0}^{N-1} \delta{t}
    \,w_1^r[J]\,S_1^r[J]^*\,Q^r[J-K]\,
    e^{i2\pi f^h_2 K\,\delta{t}}w_2^r[K]\,S_2^{hr}[K]
\end{eqnarray}
(The factors of $\delta{t}$ are to facilitate comparison with the
continuous-time idealization.)  The optimal filter $Q[J-K]$ depends
only on the difference between the two indices, consistent with the
assumption that we're considering \emph{stationary} random processes,
but the introduction of the two $N$-point window
functions\footnote{The analysis of windowing effects described here is
  a generalization of that described in \cite{windowing} and used in
  \cite{StochS1} for two data streams sampled at the same rate.}
$w_{1,2}[J]$ (assumed to be real) allows us to control edge effects by
smoothing out the onset and ending of the analyzed data, and the
complex exponential accounts for the heterodyning.

On the other hand, the analysis is not actually done with the
upsampled time series $S_1^r[J]$ and $S_2^{hr}[K]$, but with the
original $S_1[j]$ and $S_2^h[k]$.  So if we calculated a time-domain
cross-correlation statistic, it would be
\begin{equation}
  \label{eq:Ysum-tdomain}
  Y = \sum_{j=0}^{N_1-1} \delta{t}_1 \sum_{k=0}^{N_2-1} \delta{t}_2
  \,w_1[j]\,S_1[j]^*\,Q^h[j,k]\,
  w_2[k]\,S_2^{h}[k]
\end{equation}
Again, the bandpass tells us that the upsampled data streams don't
have any higher-frequency content
that's not present in the original
ones.  So we should get roughly the same cross-correlation statistic
if we limit the sum in \eqref{eq:Yrsum} to $J$ an integer multiple of
$r_1$ and $K$ an integer multiple of $r_2$, and then multiply by
$r_1r_2$ to compensate for having taken fewer terms.  This means
$Y^r\approx Y$ with
\begin{equation}
  Q^h[j,k] = Q^r[r_1j-r_2k]\,e^{i2\pi f^h_2 k\,\delta{t}_2}
\end{equation}

Since the sums over $j$ and $k$ in \eqref{eq:Ysum-tdomain} both range
from $0$ to $N-1$, the argument of $Q[j-k]$ ranges from $-(N-1)$ to
$N-1$, so a discrete Fourier transform (DFT) of $Q$ will need to
include at least $2N-1$ points.  Since it is often more convenient to
work with a $2N$-point DFT than a $2N-1$-point one (e.g., if $N$ is a
power of two or a product of small primes), we will in general
zero-pad $Q[m]$ out to $M\ge 2N-1$ points, with the ``extra'' values
(i.e., those with $N-1<m\le M-1$) defined by
\begin{equation}
  Q[m] = 
  \cases{
    0 & $N-1 < m < M-(N-1)$ \\
    Q[m-M] & $M -(N-1) \le m < M$
  }
  \ ,
\end{equation}
before defining the discrete Fourier transform
\begin{equation}
  \label{eq:Qdft}
  \Qhat[\ell] = \sum_{m=0}^{M-1} e^{-i2\pi m\ell/M}\,Q^r[m]
  = \sum_{m=-(N-1)}^{N-1} e^{-i2\pi m\ell/M}\,Q^r[m]
  \ .
\end{equation}
We can transform \eqref{eq:Ysum-tdomain} into the frequency domain
using the inverses of \eqref{eq:cft} and \eqref{eq:Qdft}:
\begin{equation}
  S_{1,2}(t)
  = \int_{-\infty}^{\infty}df\,e^{i2\pi f(t-t_0)}\, \widetilde{S}_{1,2}(f)
\end{equation}
\begin{equation}
  Q^r[m] = \frac{1}{M}\sum_{\ell=0}^{M-1} e^{i2\pi m\ell/M}\,\Qhat[\ell]
  \ 
\end{equation}
the result is 
\begin{eqnarray}
  \label{eq:Ysum-fdomain}
  Y = \frac{1}{M} \sum_{\ell=0}^{M-1} && (\delta{t})^2\,\Qhat[\ell]
  \,
  \left(\int_{-\infty}^{\infty} df_1\,
    \What_1([f_\ell-f_1]T)\,\widetilde{S}_1(f_1)\right)^*
  \nonumber
  \\
  &&\times \left(\int_{-\infty}^{\infty} df_2\,
  \What_2([f_\ell-f_2]T)\,\widetilde{S}_2(f_2)\right)
\end{eqnarray}
where $\deltaf=(M\,\delta{t})^{-1}$, $f_\ell=\ell\,\deltaf$, and the
transformed window
\begin{equation}
  \label{eq:What-def}
  \What_{1,2}(x) = \sum_{j=0}^{N-1} e^{-i2\pi xj/N_{1,2}}\,w_{1,2}[j]
\end{equation}
is equivalent to an $N_1$ or $N_2$-point discrete Fourier transform,
but not limited to integer arguments.
Note that by construction $\What_{1,2}(x)$ is periodic with period
$N_{1,2}$: $\What_{1,2}(x+N_{1,2})=\What_{1,2}(x)$.

\subsection{Mean and Variance of the Statistic}

We can get expressions for the expected mean and variance of
\eqref{eq:Ysum-fdomain} by applying \eqref{eq:s1s2corr} and
\eqref{eq:s1s1corr}, and noting that if we define $c_{ij}(f)$ by
\begin{equation}
  \langle \widetilde{s}_{i}(f)^* \widetilde{s}_{j}(f')\rangle 
  = \delta(f-f')c_{ij}(f)
\end{equation}
so that $c_{ii}(f)=\frac{1}{2}P_i(f)$ and
$c_{12}(f)=\frac{1}{2}P_{\text{gw}}(f)\gamma_{12}(f)$, then
\begin{equation}
  \langle \widetilde{S}_{i}(f)^* \widetilde{S}_{j}(f')\rangle 
  = \delta(f-f')C_{ij}(f)
\end{equation}
where
\begin{equation}
  C_{ij}(f) =
  \cases{
    c_{ij}(f) & $f^h_2 - \frac{1}{2\delta{t}_2} < f < 
    \min\left(\frac{1}{2\delta{t}_1},f^h_2 + \frac{1}{2\delta{t}_2}\right)$
    \\
    0 & otherwise
  }
\end{equation}
The mean is
\begin{eqnarray}
  \mu = \langle Y\rangle &=&
  \frac{1}{M} \sum_{\ell=0}^{M-1}(\delta{t}_1)(\delta{t}_2)\,\Qhat[\ell]
  \nonumber
  \\
  \label{eq:muexact}
  &&\times
  \int_{-\infty}^{\infty} df\,
  \What_1([f_\ell-f]T)^*\,
  \What_2([f_\ell-f]T)\,C_{12}(f)
\end{eqnarray}
As before, the restricted support of $C_{12}(f)$ means that we can
change the limits of the frequency integral from $(-\infty,\infty)$
to $(-\frac{1}{2\,\delta{t}},\frac{1}{2\,\delta{t}})$.

In this case, the periodicity of the windowing functions means that if
they're sufficiently sharply peaked about zero argument, we must have
$(f_\ell-f)T$ approximately equal to $0\mod N_1$ and $0\mod N_2$.  But
since $N$ is the lowest common multiple of $N_1$ and $N_2$, this is
equivalent to the condition that $(f_\ell-f)T\approx 0\mod N$, which
is true for at most one frequency in the range $f^h_2 -
\frac{1}{2\delta{t}_2} < f < \min\left(\frac{1}{2\delta{t}_1},f^h_2 +
  \frac{1}{2\delta{t}_2}\right)$, and that frequency is always the
positive $f_\ell$.  The windowing thus allows us to replace $C_{12}(f)$ with
$C_{12}(f_\ell)$ in \eqref{eq:muexact} and obtain
\begin{eqnarray}
  \mu &\approx&
  \frac{1}{M} \sum_{\ell=\ell_{\text{min}}}^{\ell_{\text{min}}}
  (\delta{t}_1)(\delta{t}_2)\,\Qhat[\ell]
  \,c_{12}(f_\ell)
  \nonumber
  \\
  &&\times
  \int_{-1/(2\,\delta{t})}^{1/(2\,\delta{t})} df\,
  \What_1([f_\ell-f]T)^*\,\What_2([f_\ell-f]T)
\end{eqnarray}
where the limits on the sum over $\ell$ are those such that
\begin{equation}
  f^h_2 - \frac{1}{2\delta{t}_2} < f_\ell
  < \min\left(\frac{1}{2\delta{t}_1},f^h_2 + \frac{1}{2\delta{t}_2}\right)
\end{equation}

The integral can be evaluated [using \eqref{eq:What-def}] as
\begin{eqnarray}
  &&
  \int_{-1/(2\,\delta{t})}^{1/(2\,\delta{t})} df\,
  \What_1([f_\ell-f]T)^*\,\What_2([f_\ell-f]T)
  \nonumber
  \\
  &=& \sum_{j=0}^{N_1-1} \sum_{k=0}^{N_2-1}
  w_1[j] w_2[k]
  \int_{-1/(2\,\delta{t})}^{1/(2\,\delta{t})} df\,
  e^{i2\pi (f_\ell-f)(jr_1-kr_2)\delta{t}}
  \\
  \nonumber
  & =& \frac{1}{\delta{t}} \sum_{j=0}^{N_1-1} \sum_{k=0}^{N_2-1}
  \delta_{jr_1,kr_2} w_1[j] w_2[k]
  = \frac{1}{\delta{t}} \frac{N}{r_1r_2}\overline{w_1w_2}
\end{eqnarray}
Note that the average of the product of the windows is taken only over
the points for which both windows ``co\"{e}xist'' given their
different sampling rates:
\begin{equation}
  \overline{w_1w_2} = \frac{r_1r_2}{N}
  \sum_{n=0}^{N/(r_1r_2)-1} w_1[nr_2]w_2[nr_1]
\end{equation}
This then tells us
\begin{equation}
  \label{eq:mugood}
  \mu \approx \overline{w_1w_2}\,\frac{T}{2}\,
  \sum_{\ell=\ell_{\text{min}}}^{\ell_{\text{max}}}
  \deltaf\,(\delta{t}\,\Qhat[\ell])\,\gamma(f_\ell)\,P_{\text{gw}}(f_\ell)
\end{equation}
where we have used again the definition $\deltaf=1/(M\,\delta{t})$.

As before, we can identify \eqref{eq:mugood} (up to the factor of
$\overline{w_1w_2}$) as a discrete approximation to the usual
continuous-time expression if we note that \eqref{eq:Qdft} relates the
discrete and continuous Fourier transforms according to
$
  \delta{t}\,\Qhat[\ell] \sim \widetilde{Q}(f_\ell)
$.

Note that if we design the filter in the frequency domain and chose
$\Qhat[\ell]$ to be real, \eqref{eq:mugood} tells us that the mean
value of the statistic is real as long as any underlying correlation
between the two data streams is time-symmetric.  However, the
band-passing means that $s_2(t)$ is complex, and therefore the
statistic $Y$ is as well.

Since $Y$ is complex, we consider the variance of $x := \Real Y =
\frac{Y+Y^*}{2}$ and $y := \Imag Y = \frac{Y-Y^*}{2}$ separately.  This
ultimately comes down to calculating $\langle Y^2\rangle$ and $\langle
YY^*\rangle$, and specifically the contribution from the
auto-correlations.

In fact, the dominant contribution to $\langle Y^2\rangle$ vanishes
because it contains
$
  \langle \widetilde{S}_2(f')\widetilde{S}_2(f)\rangle
$.
Now, this is zero unless both $f$ and $f'$ lie in the range $[f^h_2 -
\frac{1}{2\delta{t}_2},f^h_2 + \frac{1}{2\delta{t}_2})$, which in
particular means both $f$ and $f'$ are positive.  If it lies in that
range, it is equal to
\begin{equation}
  \langle \widetilde{s}_2(f')\widetilde{s}_2(f)\rangle =
  \langle \widetilde{s}_2(-f')^*\widetilde{s}_2(f)\rangle =
  \delta(f+f') \frac{P_2(\abs{f})}{2}
\end{equation}
but this vanishes unless $f$ and $f'$ have the opposite sign.  Thus
there is no combination of frequencies for which $\langle
\widetilde{S}_2(f')\widetilde{S}_2(f)\rangle$ is non-zero.

This then means that the real and imaginary parts of $Y$ both have a
variance of
\begin{equation}
  \sigma_x^2 \approx \sigma_y^2 = \frac{1}{2}\langle Y^* Y\rangle
  \approx
  \frac{T}{8}\,\overline{w_1^2w_2^2}\sum_{\ell=0}^{M-1}\deltaf\,
  \abs{(\delta{t}\,\Qhat[\ell])}^2\,
  P_1(\abs{f_\ell})\, P_2(\abs{f_\ell})
\end{equation}
where again the window average is conducted over the $N/(r_1r_2)$
points where the windows ``line up''.

\section{Conclusions}
We have described the data analysis method used for cross-correlating
in the frequency domain two data streams sampled at different rates,
one of which is also heterodyned prior to digitization.  We have
illustrated heuristically how this approximates a continuous-time
description, and also its equivalence to a hypothetical
cross-correlation in the time domain using upsampled data streams.
This frequency-domain technique allows us to efficiently
cross-correlate data from different detectors with different sampling
parameters, and is being applied to LIGO and ALLEGRO data taken during
LIGO's second science run.

\ack

We would like to thank everyone at the LIGO and ALLEGRO projects and
especially the LSC's stochastic analysis group.  JTW gratefully
acknowledges the University of Texas at Brownsville and the Albert
Einstein Institute in Golm.  ISH gratefully acknowledges Louisiana
State University and Albert Einstein Institute in Hannover.
Additional thanks to Lucio Baggio.
This work was supported by the National Science Foundation under
grants PHY-0140369 (Loyola),
PHY-0300609 (Loyola), PHY-0355372 (Loyola), PHY-0239735 (WSU),
and PHY-9970742 (LSU) and by the Max-Planck-Society.

\end{document}